# Anomalous Flux Pinning in $\beta$-Pyrochlore Oxide Superconductor $KOs_2O_6$

Zenji Hiroi and Shigeki Yonezawa

*Institute for Solid State Physics, University of Tokyo, Kashiwa, Chiba 277-8581*

The superconducting transition of the $\beta$-pyrochlore oxide $KOs_2O_6$ with $T_c$ = 9.60 K is studied by resistivity measurements under various magnetic fields using a high-quality single crystal. The reentrant behavior of superconductivity is observed near $T_c$ in low magnetic fields below 2 T. The recovered resistance probably due to flux flow exhibits an anomalous angle dependence, indicating that flux pinning is enhanced in magnetic fields along certain crystallographic directions such as [110], [001] and [112]. It is suggested that there is an intrinsic pinning mechanism coming from the specific crystal structure of the $\beta$-pyrochlore oxide.

$\beta$-pyrochlore osmium oxides AOs$_2$O$_6$ with A = Cs,[1)] Rb[2, 3)] and K[4)] exhibit superconducting transitions at $T_c$ = 3.3 K, 6.3 K and 9.6 K, respectively, much higher than $T_c$ = 1.0 K for the $\alpha$-pyrochlore rhenium oxide Cd$_2$Re$_2$O$_7$.[5)] They crystallize in the cubic $\beta$-pyrochlore structure with $a$ ~ 10 Å, where OsO$_6$ octahedra form a three-dimensional (3D) network involving large atomic cages for the alkali metal atoms. Band structure calculations have revealed that the electronic structure near the Fermi energy is dominated by strongly hybridized Os 5$d$ and O 2$p$ states with A states appearing far above.[6, 7)] This means that the superconducting carriers mostly reside in the 3D skeleton. A pair of Fermi spheres that are deformed octahedrally are found around the zone center. Thus, the $\beta$-pyrochlore oxides are really 3D superconductors from both structural and electronic points of view.

In the three members, KOs$_2$O$_6$ with the highest $T_c$ is distinguished from the others, exhibiting various unconventional features. For example, a jump in specific heat at $T_c$ is very large in KOs$_2$O$_6$, $\Delta C/T_c$ = 185.4 mJK$^{-2}$mol$^{-1}$,[8)] compared with the others; $\Delta C/T_c$ ~ 40 mJK$^{-2}$mol$^{-1}$ and 35 mJK$^{-2}$mol$^{-1}$ for RbOs$_2$O$_6$[9, 10)] and $\Delta C/T_c$ ~ 26 mJK$^{-2}$mol$^{-1}$ for CsOs$_2$O$_6$.[9)] The Sommerfeld coefficient $\gamma$ is estimated to be approximately 74 mJK$^{-2}$mol$^{-1}$ for the K compound[11)] and 40 mJK$^{-2}$mol$^{-1}$ for the others.[9)] Thus, $\Delta C/\gamma T_c$ is much larger than 1.4 in the former while smaller in the latter, which means that the former lies in the strong-coupling regime, while the latter in the weak-coupling regime. It has been suggested that many exceptional features on KOs$_2$O$_6$ are related to the large rattling of the K cation in an oversized cage that forms a 3D skeleton for superconducting carriers.[7, 8, 11)]

Recently, the upper critical field $H_{c2}$ of KOs$_2$O$_6$ has been determined to be 30 T by measuring rf impedance using a tunnel diode oscillator in magnetic fields up to 40 T.[12)]

The derived coherence length $\xi$ is 33 Å, which is much smaller than the penetration depth $\lambda$ = 2700 Å determined by $\mu$SR measurements.[13] Thus, the Ginzburg-Landau parameter $\kappa$ is 82, implying a typical type-II superconductor. Resistivity measurements using a high-quality single crystal indicated a residual resistivity of approximately 1 $\mu\Omega$cm, which means a large mean free path of carriers, say, 1 $\mu$m.[11] Therefore, the superconductivity of $KOs_2O_6$ lies in the clean limit, where one expects a "clean" physics in the mixed state of a type-II superconductor.

Vortex physics in a type-II superconductor has been renewed after the discovery of high-temperature superconductors (HTSCs) in cupric oxides.[14] The strong two-dimensional (2D) nature as well as large thermal fluctuations give rise to many unconventional features in HTSCs. In a classical superconductor, there is a well-defined boundary between the normal and superconducting states, that is, the upper critical field $H_{c2}(T)$ line. In HTSCs, however, the $H_{c2}(T)$ line is made obscure and instead "irreversibility line" $H_{irr}(T)$ appears, where magnetic response becomes reversible or a transition to a zero-resistive state takes place. The region between the two lines in the $H$-$T$ plane is ascribed to a vortex liquid that is caused by easy flux flow due to weak pinning. It is naively considered that the present superconductor may belong to the classical type because of its apparent 3D nature and rather low $T_c$.

Here we report an interesting flux pinning phenomenon in $KOs_2O_6$. Reentrant behavior is found in resistivity measurements under magnetic fields. A remarkable angle dependence of flux pinning is observed, suggesting that there is an intrinsic pinning mechanism ascribed to the channel structures of the $\beta$-pyrochlore oxides.

The single crystal studied in the present paper is the same as that used in our previous study, for which we reported on the "rattling phase transition".[11] The crystal shows a sharp superconducting transition at $T_c$ = 9.60 K with $\Delta T_c$ = 0.14 K in specific heat measurements. Moreover, another sharp anomaly is found at $T_p$ = 7.5 K, which may indicate a first-order phase transition associated with the rattling of the K cations. The crystal possesses a truncated octahedral shape with large (111) facets and is approximately 1.0 x 0.7 x 0.3 mm$^3$ in size and 1.3 mg in weight. Resistivity measurements were carried out by the four-probe method in magnetic fields up to 4 T in a Quantum Design Physical Property Measurement System. The crystal was rotated in magnetic fields always perpendicular to a current flow along the [1-10] direction. A typical magnitude of current was 3 mA, which corresponds approximately to a current density of 1.5 Acm$^{-2}$.

First of all, resistivity data measured at magnetic fields parallel to the [111] direction is presented. Figure 1(a) shows a set of resistivity data measured at various magnetic fields up to 4 T. At zero field, the resistivity starts to decrease from 9.7 K and reaches zero at $T_{c0}$ = 9.6 K, which coincides with the mean-field $T_c$ defined at a midpoint of the specific heat jump.[11] As magnetic field increases, the drop in resistivity shifts systematically to lower temperatures without showing a resistive broadening such as found in HTSCs.[15] However, a characteristic feature develops near the offset of the drop: tailing occurs at 0.3 T and 0.5 T, and a dip appears at 1 T. Then, at 1.5 T (blue line) the resistivity once becomes nearly zero at $T_{c0}$ = 9.1 K, recovers at 9.0 K and increases to about 10% of the normal state value, followed by a gradual decrease again to zero around 8.2 K. No sudden change in resistivity is detected near the second zero-resistive transition, in contrast to the case of YBa$_2$Cu$_3$O$_{7-\delta}$ where a drop is observed due to the vortex melting transition.[16] Such reentrant behavior is observed in a narrow field range and disappears completely above 2 T.

Figure 1(b) shows a set of isothermal resistivity curves obtained at temperatures slightly below $T_c$ as a function of magnetic field. At 9.2 K, the resistivity starts to increase from zero at $H \sim 0.15$ T, exhibits a broad hump and rapidly rises up at $H_{c2} \sim 1.1$ T toward the normal state value. As temperature decreases, the hump shifts to higher fields and exhibits zero resistance before the sharp rise. Note that at 8.8 K a reentrnat transition takes place rather sharp at 2 T. Such reentrant behavior is gone at low temperatures below 8.2 K. It is found that the finite-resistance state observed at $T < T_{c0}$ or $H < H_{c2}$ critically depends on the magnitude of current densities. For example, at $T = 9.0$ K and $H = 1.0$ T, the resistivity is nearly zero at low current density below 0.25 Acm$^{-2}$. Then, it rises up gradually with increasing current density. All the measurements shown in Fig. 1 were carried out at a current density of 1.5 Acm$^{-2}$. This fact strongly suggests that the finite resistance is due to flux flow by the Lorentz force in the present experimental geometry of magnetic field perpendicular to current. If so, it may convey a profound message on the mechanism of pinning in KOs$_2$O$_6$ that the zero-resistance state appears near $H_{c2}$, resulting in the reentrant behavior.

Figure 2 shows a summary of the above results in an $H$-$T$ phase diagram. The mean-field $T_c$ decreases linearly with increasing field, while the zero-resistive $T_{c0}$ exhibits an upward curvature first and then increases almost linearly. They coincide at zero field, but the latter is lower slightly by $\sim 0.1$ K than the former at high fields, giving rise to a narrow striped region between the two lines. The second zero-resistive line, which may correspond to the irreversibility line $H_{irr}(T)$, exhibits a strange shape. Particularly, it breaks into a narrow gap between the two finite-resistive states above and below the $H_{c2}$ line. This will be discussed later.

Anisotropy in resistivity is shown in Fig. 3. Magnetic fields are applied along the three major directions, [111], [110], and [001], of the cubic crystal. There are no detectable anisotropy in the normal state resistivity and a very small one in the zero-resistive $T_{c0}$: the difference is within 0.02 K at 3 T. It was found that even at 14 T the difference in $T_{c0}$ is at most 0.13 K, indicating really the 3D nature of the normal as well as the superconducting state of $KOs_2O_6$. In contrast, the hump below $T_{c0}$, which is observed only for the 1 T data, is apparently affected by the field direction: it appears under magnetic fields along the [111] and [001] directions, while it is completely absent in the [110] field.

In order to check the field direction dependence of the hump, the crystal was rotated at several points in the finite-resistive region of the *H-T* phase diagram shown in Fig. 2. A typical angle dependence of resistivity at $H$ = 1.0 T and $T$ = 9.0 K (marked with a cross in Fig. 2) is shown in Fig. 4. Other data taken at (*H*/T, *T*/K) = (0.75, 9.2), (1.25, 9.0) and (1.5, 8.0) exhibit essentially the same angle dependence. As the angle $\theta$ increases from zero (*H* // [111]) for [-1-11], the resistivity shows three distinct valleys with finite resistivity at $\theta$ = 19.5°, 51.5° and 89.5°. The first minimum angle is close to $\theta$ = 19.4° for the [112] direction of the crystal, and the third one to $\theta$ = 90.0° for the [-1-12] direction. In contrast, the second one is near but deviated from $\theta$ = 54.7° for the [001] direction. Moreover, there is another tiny dip at $\theta$ = 57.5°. In the reverse rotation, on the other hand, the resistivity becomes zero in a wide angle range from -30° to -46°, which is around the [110] direction with $\theta$ = -35.3°. It is to be noted that the angle dependence is not perfectly symmetrical: for example, the shapes of the peaks or valleys near [111] and [-1-11] or [112] and [-1-12], which should be identical to each other, are significantly different. It is considered that this is caused by misalignment of the crystal in the experimental setup, because we could decide precisely the [111] direction from the facet of the crystal, but the in-plane direction only approximately, so that the rotation axis of the crystal might deviate slightly from [1-10]. Thus, the deviation of angle may be enhanced with the rotation of the crystal from [111]. In fact, we carried out several experiments in the "same" geometry and found that the asymmetry changed slightly from time to time. In one experiment the second valley happened to exist exactly at [001] without an additional dip as observed in Fig. 4. This fact means that the resistivity is unusually sensitive to the field direction, especially along the [001] direction.

The observed peculiar angle dependence of resistivity suggests that flux pinning enhances along certain crystallographic directions. Although the resistivity values with $H$ // [112] and [001] are finite, they would become zero with reduced current density for measurements or other $H$-$T$ conditions. It is important to note that the shapes of the resistivity minima differ significantly among the three directions. It exhibits a flat bottom for [110], while a sharp crevasse for [001], and an intermediate valley for [112]. The flat bottom for [110] must indicate that there is a critical angle $\theta_c$ above which flux lines become free to move. $\theta_c$ is approximately 8°, although it is not easy to determine from the data in Fig. 4 because of the asymmetry of the shape. This reminds us of the lock-in transition observed in 2D superconductors such as copper oxides and layered organic conductors.[17]

Two interesting aspects are to be noted in the present study. One is the reentrant behavior and the resulting finite-resistive region in the $H$-$T$ plane at $H$ // [111]. This means that flux pinning in $KOs_2O_6$ is weak at $H$ // [111] in spite of its 3D nature and low $T_c$. In general, since $H_{irr}$ rises as $(T_c - T)^n$ with $1 < n < 2$, while $H_{c2}$ rises only as $(T_c$

– $T$), $H_{irr}$ tends to rise more rapidly and to approach $H_{c2}$, as we go further below $T_c$.[14] In the present case, however, the $H_{irr}$ line exhibits a strange shape: it turns around and tends to merge with the $H_{c2}$ line toward $T_c$. This suggests that there is an unusual pinning mechanism that is enhanced near $T_c$, not at low temperatures well below $T_c$ as in a conventional superconductor. The origin should be clarified in a future study.

The second important feature is the large sensitivity of flux-flow resistance to the field direction. As shown in Fig. 4, flux pinning may be enhanced at magnetic fields parallel to certain crystallographic directions. Note that these directions have nothing to do with the specific crystal shape, as shown in the inset of Fig. 4. Thus, the possibility of the geometrical barrier effect observed in HTSCs[18] may be excluded. It is known in type-II superconductors that extended defects in a crystal, such as dislocations, grain boundaries, inclusions of second phases and twin planes, tend to pin flux lines to sustain the Lorentz force. However, one may not expect the presence of such defects in the present high-quality crystal. Moreover, correlated defects like twin planes in $YBa_2Cu_3O_{7-\delta}$ should not exist here, because the crystal remains cubic down to low temperatures below $T_c$.[19] Even if there is a certain correlated defect, it would be difficult to explain the observed, complicated angle dependence of resistivity by assuming one type of correlated defect. Therefore, we will assume an intrinsic pinning mechanism coming from the specific crystal structure of the compound in analogy with the intrinsic pinning in HTSCs.

The superconducting carriers in $KOs_2O_6$ are located on the Os-O lattice. The Os sublattice comprises the pyrochlore lattice that contains large atomic cages or channels when viewed along a certain crystallographic direction. The largest channel appears along [110], as shown in Fig. 5(a). The size of the channel is about 5 Å in diameter.

As the crystal is rotated for [001], smaller channels appear at [112] and [001]. Other projections always give such a dense array of atoms as in [111] shown in Fig. 5(d). It is considered that flux pinning is so strong at $H$ // [110] that flux lines seem to be trapped strictly along the channels, remaining locked in below a finite lock-in angle $\theta_c$. This is the scenario proposed for HTSCs with very small coherence lengths along the $c$ axis, say $\xi \sim 3$ Å, compared with a separation at least 6 Å between two superconducting $CuO_2$ planes.[14] In contrast, the coherence length of $KOs_2O_6$ is large, $\xi = 33$ Å, much larger than the size of the channels. Nevertheless, we believe that the channel structures are the only possible solution to interpret the present experimental results in a reasonable way. Presumably, the channels could be the source of collective pinning for a rather rigid flux line lattice.[20]

In summary, we reported on the anomalous flux pinning in the β-pyrochlore superconductor $KOs_2O_6$. It was demonstrated that the magnitude of flux pinning dramatically changes with the field orientation; large at $H$ // [110], [112], and [001]. In contrast, pinning is weak at $H$ // [111], which results in an unusual shape for the irreversibility line in the $H$-$T$ plane. An intrinsic pinning mechanism coming from the channel structure of the compound was proposed.

**Acknowledgements**

We are grateful to T. Tamegai for helpful discussions. This research was supported by a Grant-in-Aid for Scientific Research B (16340101) and Scientific Research on Priority Areas "Invention of Anomalous Quantum Materials" provided by the Ministry of Education, Culture, Sports, Science and Technology, Japan.

Figure caption

Fig. 1. (Color online)  (a) Evolution of resistivity curves measured on cooling in various magnetic fields parallel to the [111] direction.  The magnetic fields from the right to the left curves are 0, 0.3, 0.5, 1.0, 1.5, 1.75, 2.0, 3.0 and 4.0 T.  The upper panel shows an enlargement around the offset of the resistivity drop.  (b) Set of isothermal resistivity curves as a function of magnetic field.  The temperatures are 9.2, 9.0, 8.8, 8.6 and 8.0 K from the left to the right curves.

Fig. 2. (Color online)  $H$-$T$ phase diagram for $KOs_2O_6$ in magnetic fields parallel to the [111] direction.  Red circles represent the mean-field $T_c$ determined by the previous specific heat measurements,[11] and blue diamonds the zero-resistive $T_{c0}$ determined in the present resistivity measurements.  Error bars on the red circles indicate a 10-90% transition width in specific heat.  The region surrounded by green triangles and squares represents a finite-resistivity region in the superconducting state.  The triangles and squares are obtained by constant-field and isothermal scans, respectively.  The region colored in yellow gives a zero-resistive state.

Fig. 3. (Color online)  Resistivity measured in magnetic fields parallel to the [111], [001] and [110] directions.  A hump near the offset of the supreconducting transition is observed in the [111] and  [001] directions and $H = 1$ T, not in the [110] direction nor in other fields.  The anisotropy in both the normal-state resistivity and $T_c$ is very small.

Fig. 4. (Color online)  Angle dependence of resistivity measured at $H = 1.0$ T and $T = 9.0$ K, whose position is indicated by cross in Fig. 2.  The current flow is along [1-10], which is always perpendicular to applied fields.  The approximate shape of the crystal viewed along the current direction is depicted with typical field directions by arrows.  Zero angle means the [111] direction.

Fig. 5. (Color online) Crystal structure of the *β*-pyrochlore oxide viewed along four crystallographic directions; (a) [110], (b) [001], (c) [112] and (d) [111]. Only osmium atoms are shown, which form the pyrochlore lattice and must contain most of superconducting carriers. The dotted line represents the projected unit cell with *a* = 10.1065 Å.

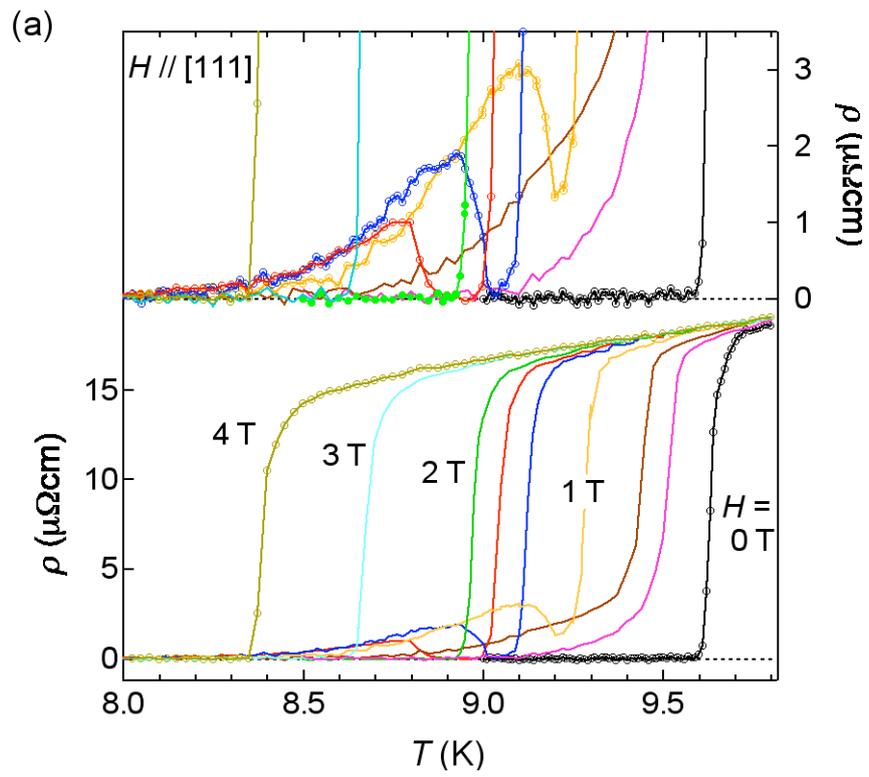

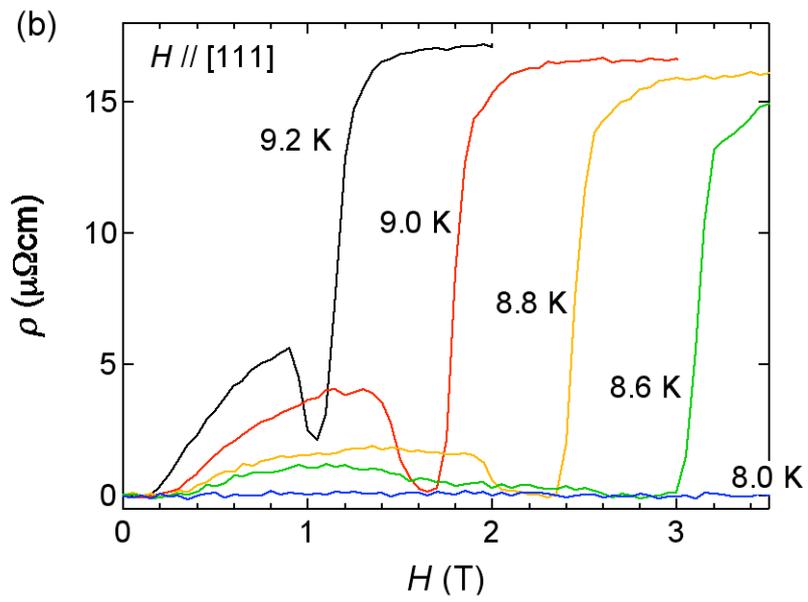

Fig. 1

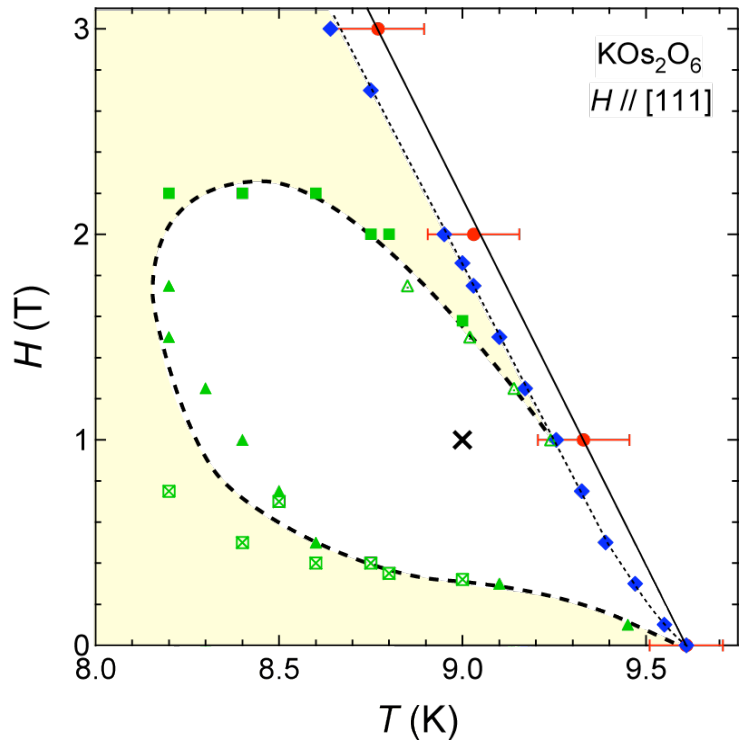

Fig. 2

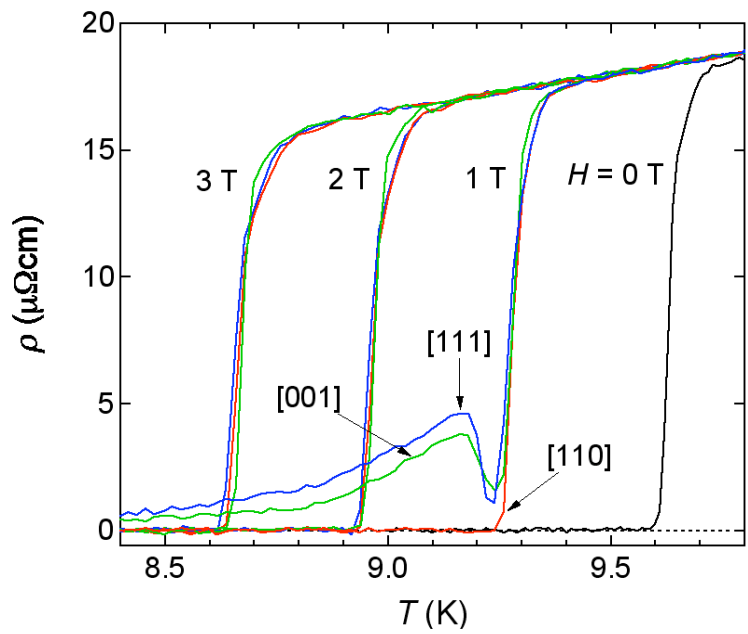

Fig. 3

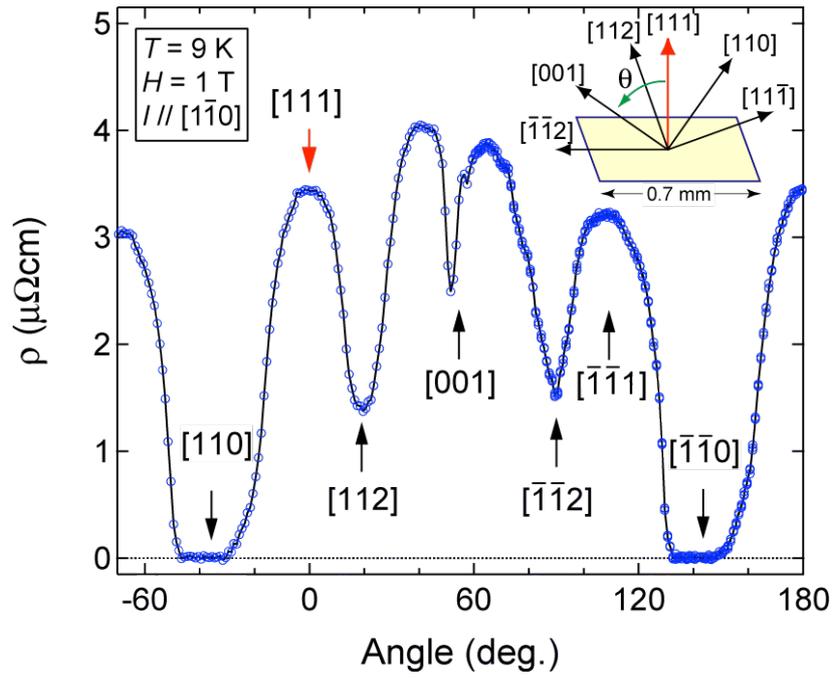

Fig. 4

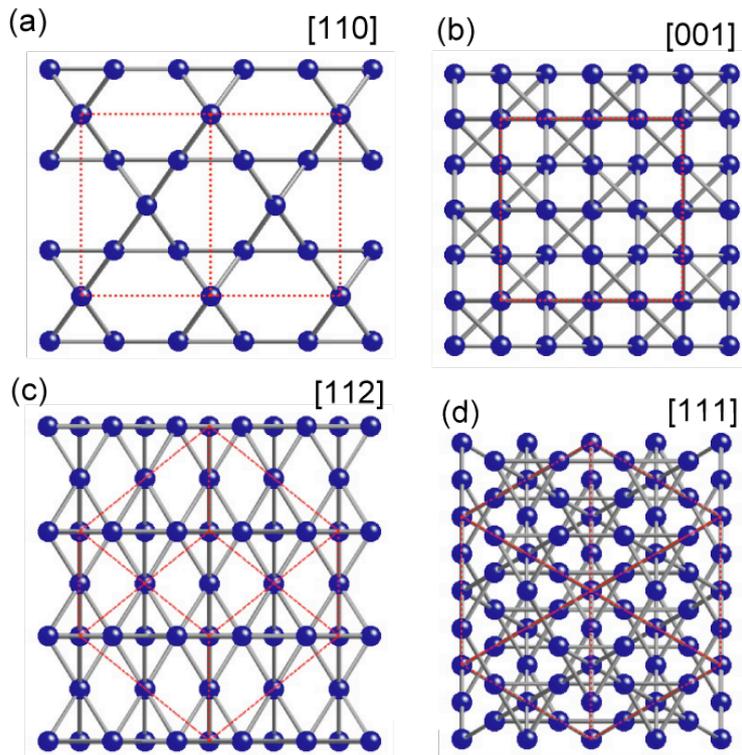

Fig. 5

15